\documentclass[12]{article}
\usepackage{graphicx}

\begin{document}

\title {THE POLARIZATION EFFECTS OF RADIATION FROM MAGNETIZED ENVELOPES AND EXTENDED ACCRETION STRUCTURES}

\author {Gnedin\, Yu.N.$^1$, Silant'ev\, N.A.$^{1,2}$
\\Piotrovich\, M.Yu.$^1$, Pogodin\, M.A.$^1$}

\maketitle

\begin{center}
{\small (1) Central Astronomical Observatory at Pulkovo,
Saint-Petersburg, Russia.

(2) Instituto National de Astrofisica, Optica y Electronica,
Apartado Postal 51 y 216, C.P. 72000, Pue. Mexico.}
\end{center}

\begin{abstract}
The results of numerical calculations of linear polarization from
magnetized spherical optically thick and optically thin envelopes
are presented. We give the methods how to distinguish magnetized
optically thin envelopes from optically thick ones using observed
spectral distributions of the polarization degree and the
positional angle. The results of numerical calculations are used
for analysis of polarimetric observations of OB and WR stars,
X-ray binaries with black hole candidates (Cyg X-1, SS 433) and
supernovae. The developed method allows to estimate magnetic field
strength for the objects mentioned above.
\end{abstract}

\section{INTRODUCTION}

The existence of magnetic fields in hot stellar atmospheres and
envelopes as well as in accretion envelopes of quasars and AGNs
gives rise to the Faraday rotation of the polarization plane of
the radiation from these objects. The angle of the Faraday
rotation $\psi$ may be written in the form [1]:

\begin{equation}
\psi({\bf n, B})=\frac{1}{2}\delta \tau_T \cos\Theta, \label{1}
\end{equation}
\begin{equation}
\delta=\frac{3}{4\pi}\cdot \frac{\lambda}{r_e}\cdot
\frac{\omega_B}{\omega}\cong 0.8\lambda^2(\mu m)B(G). \label{2}
\end{equation}

\noindent where $\bf n$ is the direction of wave propagation,
$\Theta$ is the angle between $\bf n$ and magnetic field $\bf B$,
$\tau_T$ is the Thomson optical length ($\tau_T = N_e \sigma_T
l$), $N_e$ is the electron number density, $l$ is the geometric
path, $\sigma_T = 8\pi r_e^2 / 3$ is Thomson's cross-section, $r_e
= e^2/mc^2 \approx 2.82 \times 10^{-13}$cm is the classic radius
of electron, $\omega_B = |e|B/mc$ is the cyclotron frequency,
$\omega = 2\pi c / \lambda = kc$ is the frequency of radiation
($\omega_B / \omega \approx 0.93 \times 10^{-8} \lambda(\mu
m)B(G)$).

At large distances from a star or a quasar magnetic field is
approximately dipolar because other multipoles fall off with
distance more rapidly than the dipole component. Because of small
value of Thomson's cross-section $\sigma_T$ the optically thick
envelopes seem be seldom the case. The expanding matter just after
the supernova explosion, when the concentration of particles is
yet large enough, is possibly a case of such optically thick
envelope. Another example may be the accretion structures near
quasars and AGNs.

Magnetospheres around the black holes of different masses have to
play a very important role in formation of various structures of
outflowing matter (winds, expanding coronae, jets, etc.). Possible
existence of such magnetospheres was confirmed recently both by
spectroscopic (Robertson and Leiter [2]) and by polarimetric
(Gnedin at al. [3]) observations of X-ray binary systems with
black holes.

Now the Blandford and Znajek mechanism is commonly accepted which
provides the energy extraction from a rotating black hole by means
of magnetic field [4].

Later this idea was developed in papers [5-7]. In these papers the
model was considered where the toroidal electric current at the
inner boundary of the accretion disk generates the dipole magnetic
field connecting together the black hole horizon and the region of
the accreting disk beyond its inner boundary.

The plasma outflow from the magnetosphere and the inner part of
accretion disk gives rise to an envelope. The radiation scattered
in the envelope acquires the linear polarization. The region of
the formation of broad emission lines observed in AGNs is a case
of such envelope.

In the absence of magnetic field the scattered radiation has to be
non-polarized provided the scattering matter in the spherically
shape envelope is symmetric relative to the line of sight. If the
magnetic field of the object is not symmetric relative to this
line, the outgoing radiation (singly or multiply scattered) has to
undergo influence of the Faraday's rotation effect and becomes to
be polarized. The integral polarization of radiation is equal to
zero if the dipole magnetic momentum $\bf M$ is coaxial with the
line of sight or in the case of ${\bf B}=0$.

\section{OPTICALLY THIN ENVELOPE}

The integral polarization scattered in an envelope is simplest to
calculate if the envelope is optically thin. In this case we can
take into account only single scattered radiation and the Faraday
rotation during the followed propagation to an observer. This was
made for various types of envelopes in papers [8-10], where,
however, was accepted that a star is a point-like source of
non-polarized radiation. This assumption gives rise to some
overestimation of the polarization degree (more detailed, see
below). Maximal disturbance of axial symmetry of the Faraday's
rotation picture occurs when magnetic dipole $\bf M$ is
perpendicular to line of sight $\bf n$ ($\vartheta_m =
90^{\circ}$). In this case the integral polarization from the
whole envelope acquires its maximum value and the polarization
plane coincides with the plane ($\bf n M$).

The polarization Stokes parameters $F_Q$ and $F_U$, according to
[1, 8, 9], are described by the following expressions:

\begin{equation}
F_{Q}({\bf n})=-\frac{L}{4\pi R^2}\cdot
\frac{3}{16\pi}\,\sigma_{T} \int\,dV\frac{N_e({\bf
r})}{r^2}\sqrt{1-\frac{R_s^2}{r^2}}
\sin^2\vartheta\,\cos2(\varphi+\psi), \label{3}
\end{equation}

\begin{equation}
F_{U}({\bf n})=-\frac{L}{4\pi R^2}\cdot
\frac{3}{16\pi}\,\sigma_{T} \int\,dV\frac{N_e({\bf
r})}{r^2}\sqrt{1-\frac{R_s^2}{r^2}}
\sin^2\vartheta\,\sin2(\varphi+\psi), \label{4}
\end{equation}

\noindent where $L(erg\, s^{-1}\,Hz^{-1})$ is the luminosity of
the central star, $R$ is the distance from the star to the
telescope, the angles $\vartheta$ and $\varphi$ determine the
direction of the radius vector ${\bf r}(r, \vartheta, \varphi)$;
$R_s$ is the radius of the star. Integration yields over the
visible part of the envelope volume. In (3) and (4) we accepted
that X-axis lies in the plane ($\bf n M$).

Near the stellar surface the polarization of the scattered
radiation is small because of the non-polarized star's radiation
is falling to the volume element $dV$ practically isotropic from
the solid angle $\approx 2\pi$. In the book [1](chapter 4) the
formulae are given which describes this effect quantitatively.
There was considered the commonly used model when the intensity
$I(\mu)$ of outgoing radiation is approximated by the formula
$I(\mu) = A + C\mu$. Remember that $\mu$ is the cosine of the
angle between the normal to the atmosphere's surface and the
direction of the escaping photons. In this paper we consider only
the isotropic case ($C = 0 $, Lambert's law of radiation) when the
effect of depolarization of scattered radiation near the stars's
surface is maximal (most profound). This gives rise to the factor
$\sqrt{1 - R_S^2/r^2}$ in formulae (3) and (4) as compared with
the analogous formulae in [8-10] where a star was considered as
point-like source of non-polarized radiation. The non point-like
character of the radiation source was taken into account first in
[11].

We restrict ourselves by two types of the electron number density
$N_e(\bf r)$ in an envelope: $N_e(r) = N_0 = const$ and $N_e({\bf
r}) = N_0(R_s^2/r^2)$. The explicit formulae for Faraday's
rotation angle $\psi$ have the form [8]:

For the envelope with $N_e(r) = const$

\begin{equation}
\psi =\frac{\delta_s
\tau_{env}}{2(\eta-1)\rho^2}\{\cos\vartheta_m\,
[\cos\vartheta-\frac{\rho^2}{\eta^2}\sqrt{1-\frac{\rho^2}{\eta^2}
\sin^2\vartheta}]+\sin\vartheta_m\cos\varphi\sin\vartheta(1-\frac{\rho^3}
{\eta^3})\}. \label{5}
\end{equation}

\noindent Here, $\eta=R_0/R_s$ is the ratio of the envelope radius
$R_0$ to the radius of a star $R_s$; $\rho=r/R_s$,
$\delta_s=0.8\lambda^2 M/R_s^3$ is the Faraday's depolarization
parameter (2) at the magnetic equator of the star, $M$ is magnetic
dipole moment of the star, $\tau_{env}=N_0\sigma_T (R_0-R_s)$ is
Thomson's optical depth of the envelope, $\vartheta_m $ is the
angle between $\bf M$ and line of sight $\bf n$.

For the envelope with $N_e = N_0(R_s^2/r^2)$:

\begin{equation}
\psi=\frac{\delta_s\tau_{env}}{2\rho^4 \sin^4\vartheta}
\{\cos\vartheta_m\,[-(1-\cos\vartheta)+\frac{4}{3}(1-\cos^3\vartheta)-
\frac{3}{5}(1-\cos^5\vartheta)]+\frac{3}{5}\sin{\vartheta_{m}}
\cos\varphi \sin^5\vartheta \}. \label{6}
\end{equation}
\noindent Here, $\tau_{env}=N_0\sigma_T\,R_s$ is the optical depth
of the envelope. Remind that for single scattered radiation
($\tau_{env}\ll 1$) we neglect the terms $\sim \tau^2$ and higher.
On this approximation when calculating the polarization degree
$p_{\it l}=\sqrt{F_Q^2+F_U^2}/F_I$ one can take for the flux $F_I$
the value $L/4\pi R^2$.

It follows from (3) and (4) that $F_Q$ and $F_U$ have the form:

\begin{equation}
F_{Q,U}= F_I\,\tau_{env}\,f_{Q,U}(\delta_s \tau_{env}, \vartheta_m
,\eta ). \label{7}
\end{equation}

For the envelope with $N_e\sim r^{-2}$ the variable $\eta$ in (7)
is absent.

For weak magnetic fields or large wavelengthes, when $\psi \ll 1$,
the $f_{Q,U}$ functions have the following asymptotic form:

\[
f_Q\simeq K_Q\,(\delta_s\,\tau_{env})^2\,\sin^2\vartheta_m,
\]

\begin{equation}
f_U\simeq
K_U\,(\delta_s\,\tau_{env})^3\,\cos\vartheta_m\,\sin^2\vartheta_m.
\label{8}
\end{equation}

The case $\psi \ll 1$ corresponds to $\delta_S\tau_{env} \ll 1$.
It means that $|f_U|\ll |f_Q|$, i.e. the wave electric field
oscillates near the plane ($\bf n M$) and the polarization degree
$p_l\approx \tau_{env}|f_Q|$. The asymptotic formulae (8) take
place up to $\delta_S\tau_{env}\approx 1$.

\begin{table}
\caption[]{The coefficients K$_{\rm Q}$, K$_{\rm U}$ and the
maximal polarization degree for the various values of the
parameter $\eta $ (the model of a non-point like source of
radiation).} {\tiny
\begin{flushleft}
\begin{tabular}{lllllllllll}
\hline \noalign{\smallskip} ~~$\eta $ &~~1.02 &~~1.05 &~~1.1
&~~1.2 &~~1.5 &~~2 &~~3 &~~4 &~~5 \\ \noalign{\smallskip} \hline
\noalign{\smallskip}
  ~100\,K$_{\rm Q}$    &~~0.747 &~~1.116  &~~1.362 &~~1.376 &~~0.830 &~~0.325
&~~0.078 &~~0.028 &~~0.013 \\
  ~100\,K$_{\rm U}$ & $-$2.644 & $-$2.944 & $-$2.593 & $-$1.638 & $-$0.375
\medskip
& $-$0.048 & $-$0.002 & $-$0.000 & $-3\cdot 10^{-5}$ \\
$\frac{{\rm P}_{l\,max}}{\tau_{env}}$(\%) &~~0.974 &~~1.650
&~~2.457 &~~3.566
\medskip
&~~5.252 &~~6.075 &~~6.081 &~~5.964 &~~5.841 \\ $\sqrt{\delta_s
\,\tau_{env}}$ &~~1.40 &~~1.45 &~~1.50 &~~1.65 &~~2.10 &~~2.80
&~~5.05 &~~7.60 &~10.30 \\ \noalign{\smallskip} \hline
\end{tabular}
\end{flushleft}
}
\end{table}

\begin{table}
\caption[]{The same values for the model of a point like star.}
{\tiny
\begin{flushleft}
\begin{tabular}{lllllllllll}
\hline \noalign{\smallskip} ~~$\eta $ &~~1.02 &~~1.05 &~~1.1
&~~1.2 &~~1.5 &~~2 &~~3 &~~4 &~~5 \\ \noalign{\smallskip} \hline
\noalign{\smallskip}
  ~100\,K$_{\rm Q}$    &~~~~8.018 &~~~~7.593  &~~~~6.631 &~~4.892 &~~2.065
&~~0.662 &~~0.139 &~~0.048 &~~0.022 \\
  ~100\,K$_{\rm U}$ & $-$28.32 & $-$19.67 & $-$12.27 & $-$5.592 & $-$0.869
\medskip
& $-$0.086 & $-$0.003 & $-$0.000 & $-$0.000 \\ $\frac{{\rm
P}_{l\,max}}{\tau_{env}}$(\%) &~~~8.136 &~~~8.728 &~~~9.310
\medskip
&~~9.830 &~10.03 &~~9.241 &~~7.403 &~~6.609 &~~6.234 \\
$\sqrt{\delta_s \,\tau_{env}}$ &~~~1.30 &~~~1.35 &~~~1.40 &~~1.55
&~~1.90 &~~2.55 &~~3.65 &~~6.30 &~10.00 \\ \noalign{\smallskip}
\hline
\end{tabular}
\end{flushleft}
}
\end{table}

The numerical coefficients $K_Q$ and $K_U$ for the case $N_e\sim
r^{-2}$ are equal to 0.0006641 and -0.0000796, correspondingly.
For the model of point-like source of radiation these values are
considerably larger: 0.0016136 and -0.0001873. In the case $N_e =
const$ the coefficients $K_Q$ and $K_U$ depend on the parameter
$\eta = R_0 / R_S$. The numerical values of $K_Q(\eta)$ and
$K_U(\eta)$, as functions of $\eta$, for non point-like star, are
presented in Table 1. It should be noted that for $\eta < 1.15
(K_Q(1.15) = 0.014136, K_U(1.15) = -0.02088)$ the polarization of
radiation decreases. This fact reflects the effect of
depolarization of scattered radiation near the star surface,
mentioned above. On the contrary the values of $K_Q(\eta)$ and
$K_U(\eta)$ for the point-like star model (see, Table 2) increase
monotonously with $\eta \rightarrow 1$. The negative sign of $K_U$
denotes that the positional angle $\chi$ ($\tan{2\chi} = f_U /
f_Q$) is also negative, i.e. corresponds to left-hand rotation of
the wave electric fields from the plane ($\bf n M$), for the line
of sight $\bf n$.

The values $p_l/\tau_{env} = \sqrt{(f_Q^2 + f_U^2)}$ (in \%) and
the positional angle $\chi$ (in degrees), as the functions of
$\sqrt{\delta_S\tau_{env}}$, are presented in Fig.1. Remember that
the X-axis lies in the plane ($\bf n M$). The upper panels
correspond to $N_e(r) = N_0(R_S/r)^2$ and the lower ones
correspond to envelope with $N_e = const$ and $\eta = 5$. All
curves represent the model of non-point-like star. For the
polarization degree to be obtained the right-hand curves must be
multiplies by the optical depth $\tau_{env} < 1$.

For optically thin envelopes the maximum polarization $p_{l\,
max}$ (at $\vartheta_m = 90^{\circ}$) can acquire rather
considerable values $\sim 6\tau_{env}$(\%) for the non-point-like
star (see Table 1) and $\sim 10\tau_{env}$ (\%) for point-like
model of the radiation source (see Table 2). In these tables we
present the values of $\sqrt{\delta_S\tau_{env}}$ corresponding to
$p_{l\, max}$ values.

It is notable that $p_{l\, max}$ is almost the same for various
geometric depths of the envelopes ($\eta\simeq 1.5\div 5$ for
non-point-like stars) and ($\eta\simeq 1.02\div 3$ for point-like
model). Only the corresponding values of
$\sqrt{\delta_S\tau_{env}}$ parameter change its position.
Considering the fixed value $\tau_{env}$ for all envelopes we can
interpret the mentioned fact as for every envelope there exists
the effective layer with $\psi\approx 1$ which gives the main
contribution to the integral polarization. The interval of
$\sqrt{\delta_S\tau_{env}}$, corresponding to $p_{l\, max}$,
increases with the increasing of the envelope radius $R_0$. For
$\eta = 2$ this interval (the width of the polarization spectrum)
corresponds to $\sqrt{\delta_S\tau_{env}}\approx 2\div 5$. For
$\eta = 5$ this width corresponds to
$\sqrt{\delta_S\tau_{env}}\approx 4\div 20$.

With increasing magnetic field this effective layer shifts from
the stellar surface up to the outer boundary of the envelope. For
further increasing of $\bf B$ the polarization is determined by
the thin boundary layer where $\psi\leq 1$. This condition takes
place at the optical depth of this layer $\tau\sim 1/\delta_0
\propto \lambda^{-2}$, where $\delta_0=0.8 \lambda^2 M/R^3_0$ is
the parameter (2) at the outer boundary of the envelope. This
estimation assumes that the total inner volume does not give a
contribution to integral polarization, i.e. $\delta_s
\tau_{env}/\eta^3\gg 1$. The intensity of the radiation scattered
in this thin layer is proportional to $\tau$. It means that the
integral polarization is also inversely proportional to
$\delta_0$, i.e. $p_{\it l}\simeq C(\vartheta_m)\eta^3/\lambda^2$.
The analytical formula for $\vartheta_m=90^{\circ}$ gives:

\begin{equation}
p_{\it l}\approx \frac{\pi
\eta^3}{16\delta_s}\sqrt{1-\frac{1}{\eta^2}} \propto
\frac{1}{\lambda^2}, \,\,\,\,\,\, \frac{\delta_s
\tau_{env}}{\eta^3}\gg 1. \label{9}
\end{equation}

Asymptotic dependence $p_{\it l}\propto \lambda^{-2}$ takes place
for $\delta_s \tau_{env}/\eta^3\approx 50\div 100$. The expression
(9) was obtained without the exclusion of invisible part of the
envelope, i.e. it gives rather overestimated value for $p_{l}$.
So, for $\eta = 2$ the overestimation is equal to 11\% at
$\sqrt{\delta_S\tau_{env}} = 1$ and equal to 3\% for
$\sqrt{\delta_S\tau_{env}} = 10$. For more vast envelope with
$\eta = 5$ the corresponding values are 6\% and 0.6\%. For the
model with $N_e\propto r^{-2}$ these values of overestimation are
equal to 15\% and 2\%, correspondingly. If $\eta\rightarrow 1$,
the invisible part of the envelope are practically the half of all
the envelope volume. Therefore, for this case we need to diminish
the value (9) to 2 times.

As it was showed in[1], a star can be considered as a point-like
source of non-polarized radiation from the distances $(1\div
2)R_s$ from the star surface. Therefore, the asymptotic values of
polarization at $\delta_S\tau_{env}\gg 1$ are the same for both
models of point-like and non-point-like sources, if, of course,
the geometric depth of an envelope is larger than $(2\div 3)R_S$.

For envelopes with $N_e\sim r^{-2}$ the peak of polarization (at
$\vartheta_m =90^{\circ}$) occurs at
$\sqrt{\delta_S\tau_{env}}\approx 4.65$ and equal to
$2.886\tau_{env}$(\%) for the case of non point-like model. The
point-like model for the same value
$\sqrt{\delta_S\tau_{env}}\approx 4.65$ gives
$3.878\tau_{env}$(\%). The peak polarization for point-like star
model occurs at $\sqrt{\delta_S\tau_{env}}\approx 3.05$ and is
equal to $4.722\tau_{env}$(\%).

Asymptotic behavior of polarization at
$\sqrt{\delta_S\tau_{env}}\gg 1$ and $N_e\sim r^{-2}$ is
determined also by outer layer of the envelope ($r\geq r_*$) where
$\psi (r_*)=0.5\delta (\tau_*)\tau (r_*)\leq 1$. The optical depth
of this layer is equal to $\tau (r_*)= t_{env}R_S/r_*$, where
$\delta (r_*)$ is the value of parameter (2) at this boundary
$r=r_*$. Taking into account that $\delta(r_*)\propto 1/r_*^3$, we
obtain $r_*\propto\sqrt{\lambda}$. This means $p_l\propto
1/\sqrt{\lambda}$. To calculate the polarization at large values
of the parameter $\sqrt{\delta_S\tau_{env}}$ we can, as in
previous case, take the numerical value at
$\sqrt{\delta_S\tau_{env}}= 10$ from the Fig.1 and extrapolate it,
using asymptotic dependence on $\lambda$. For $\vartheta_m
=90^{\circ}$ and $N_e\sim r^{-2}$ one yields the formula:

\begin{equation}
f_Q\approx \frac{0.086}{(\delta_s\tau_{env})^{1/4}},\propto
\frac{1}{\sqrt{\lambda }}, \,\,\,\,\,\,\delta_s\tau_{env}\gg 1.
\label{10}
\end{equation}

Thus, for $N_e\sim r^{-2}$ the polarization degree decreases as
$\sim 1/\sqrt{\lambda}$. Using qualitative derivation of
$p_l(\lambda)$ at $\sqrt{\delta_s\tau_{env}}\gg 1$, given above,
one can derive for $N_e\propto r^{-\nu}$ the following asymptotic
$\lambda$-dependence $p_l\propto \lambda^{-2(\nu -1)/(\nu +2)}$.

The polarization at $\delta_S\tau_{env}\gg 1$ is determined by the
contribution of the envelope's volumes far from the central
source. It means that both the polarization degree and the
positional angle give an information about number density $N_e(r)$
and magnetic field $\bf B(r)$ just in this part of the envelope.
So, polarimetric observations can determine (or to prove some
accepted model) the distribution of free electrons in the
envelope.

\section{OPTICALLY THICK ENVELOPE}

To calculate the integral polarization of radiation from optically
thick envelopes one has to know the intensity $I$ and the Stokes
parameters $Q$ and $U$ of the radiation outgoing from an element
of the surface of the envelope. We assume that the sources of
radiation with the considered wavelength $\lambda$ are distributed
far from the outer surface of the envelope, i.e. the parameters
$I, Q, U$ are the solution of the classic Milne's problem for
semiinfinite plane-parallel atmosphere with taking into account
the Faraday rotation of the polarization plane. This problem was
solved numerically (see, [12, 13]) only for the magnetic field
directed along the normal to the atmosphere. For the arbitrary
direction of magnetic field the problem is too complicated and is
not solved yet. Considering the dipole, i.e. non-homogeneous,
magnetic field we are to know the solution of the Milne problem
for arbitrary magnetic field direction.

In the paper [14] the simple asymptotic formulae (at $\delta_0
\geq 1$) for $I, Q, U$ are derived for a number of radiative
transfer problems, including the Milne problem. It is important
that these analytic formulae are suitable for the arbitrary
direction of magnetic field. They give exact solutions for large
values of the depolarization parameter $\delta$ (see (2)). These
formulae correspond to the approximation when the intensity of
radiation is found from the usual scalar radiation transfer
equation with the Rayleygh phase function and the polarization is
calculated as a result of single scattering of known radiation
flux before the outgoing from the semiinfinite atmosphere. The
transformation of this single scattered radiation due to Faraday's
rotation  effect is also taken into account. These formulae give
to some extent overestimated values for the polarization. The
comparison with the exact numerical results for magnetic field
directed along the normal to the atmosphere shows that for the
conservative case ($q=\sigma_a/(\sigma_a + \sigma_T)=0$) these
asymptotic formulae give the correct result with the relative
error of 10\% at $\delta_0 = 10$. For $\delta_0 = 5$ the error is
larger and is equal to $\approx 20\%$. For absorbing atmosphere
($q\neq 0$) these asymptotic formulae are less exact then for more
important case $q = 0$.

The use of these simple asymptotic formulae (see [14]) gives rise
to the following expressions for the Stokes parameters of the
radiation from an optically thick magnetized spherical envelope:

\begin{equation}
F_I=\frac{a^2}{R^2}F, \label{11}
\end{equation}

\begin{equation}
F_Q=- \frac{a^2}{R^2}\cdot \frac{F}{\pi J_1}\cdot \frac{1-g}{1+g}
\int\limits_0^1\!d\mu\!\int\limits_0^{\pi
}\!d\varphi\,\mu(1-\mu^2)\,\frac {(1-k\mu)\,\cos(2\varphi
)}{(1-k\mu)^2+[(1-q)\delta \cos\Theta]^2}, \label{12}
\end{equation}

\begin{equation}
F_U=- \frac{a^2}{R^2}\cdot \frac{F}{\pi J_1}\cdot \frac{1-g}{1+g}
\int\limits_0^1\!d\mu\!\int\limits_0^{\pi
}\!d\varphi\,\mu(1-\mu^2)\,\frac
{(1-q)\delta\cos\Theta\,\cos(2\varphi )} {(1-k\mu)^2+[(1-q)\delta
\cos\Theta]^2}. \label{13}
\end{equation}

\noindent Here, $F$ is the radiation flux from the $1\,cm^2$ of
the envelope surface, $a$ is the outer radius of the envelope, $R$
is the distance to an observer, the numerical factors $J_1$ and
$g$ depend on $q$ and are calculated in [14]. The value
$\delta\cos{\Theta}$ for the dipole magnetic field is equal to

\begin{equation}
\delta \cos\Theta =\delta_0\,[\,(3\mu^2-1)\,\cos\vartheta_m +
3\mu\sqrt{1-\mu^2}\sin\vartheta_m \cos\varphi\,], \label{14}
\end{equation}

\noindent where $\delta_0 =0.8 \lambda^2 M/a^3$ is the parameter
(2) at the magnetic equator of outer surface of the envelope.

Further, we shall consider also two cases of the distorted dipole
magnetic field: 1) diamagnetic (perfectly conducting) envelope
when the normal component of the magnetic field does not penetrate
into medium, and 2) the envelope with the strong plasma radial
outflow when the frozen magnetic field acquires the radial form
(this is the analogy of Parker's spherically symmetric wind for
non-rotating star). The envelope can be considered as diamagnetic
one when the ohmic diffusion characteristic time
($\tau_{ohm}\approx a^2/6 D_m$ with ohmic diffusion coefficient
$D_m=c^2/4\pi \sigma $) is much lesser than others characteristic
times (the time of turbulent mixing, period of cyclotron rotation
and so on (see [15])).

For diamagnetic envelope the radial component $B_r$ of the dipole
magnetic field are to be excluded. This gives the expression:

\begin{equation}
\delta \cos \Theta =\delta_0\,[\,-\cos \vartheta_m\,(1-\mu^2) +\mu
\sqrt{1-\mu^2} \sin \vartheta_m\,\cos\varphi\,]. \label{15}
\end{equation}

For the Parker outflow, on the contrary, only $B_r$ must be taken
into account:

\begin{equation}
\delta \cos \Theta =2\,\delta_0\,(\mu^2\,\cos \vartheta_m +\mu
\sqrt{1-\mu^2} \sin \vartheta_m\,\cos\varphi\,). \label{16}
\end{equation}

In Figures the cases (14), (15) and (16) are denoted as a), b) and
c).

For small values of the depolarization parameter $\delta_0 \ll 1$
and the most interesting case of conservative atmosphere ($q = 0$)
the integrals (12) and (13) can be easily calculated analytically.
In this case we have $F_Q\propto\delta_0^2$ and
$F_U\propto\delta_0^3\ll F_Q$. It gives $p_l=\sqrt{F_Q^2 +
F_U^2}/F_I$ and $\tan{2\chi}=F_U/F_Q$. Finally we obtain

\begin{equation}
p_{\it l}\simeq C\,\delta_0^2\,\sin^{2}\vartheta_m , \label{17}
\end{equation}

\begin{equation}
\chi \simeq D\,\delta_0\,\cos\vartheta_m , \label{18}
\end{equation}

\noindent where the coefficients $C$ and $D$ for the cases (14),
(15) and (16) are equal to: $C=0.765$, $D=17.19$ for the dipole
(14); $C=0.0797$, $D=-51.56$ for the diamagnetic envelope (15);
$C=0.319$, $D=68.75$ for the radial outflow (16). It should be
noted that X-axis is chosen in the plane "line of sight - magnetic
dipole" ($\bf n M$). The positive $\chi$ corresponds for an
observer to the anticlockwise deviation from this plane ($\bf n
M$). If the magnetic dipole $\bf M$ has $\vartheta =90^{\circ}$,
the sign of the positional angle $\chi$ is opposite.

Eqs. (17) and (18) show that $p_l\propto\lambda^4$ and
$\chi\propto\lambda^2$. The comparison with numerical calculations
demonstrates that eqs. (17) and (18) are valid up to $\delta_0 =
0.25$.

\section{DISCUSSIONS OF THE RESULTS}

Numerical calculations of $p_l$ and $\chi$ from (11)-(13) are
presented in Fig.2-5 as functions of dimensionless parameter
$\sqrt{\delta_0}\cong 0.89\,\lambda(\mu m)\sqrt{B_0(G)}$, where
$B_0=M/a^3$ is the magnetic field on the surface of optically
thick spherical envelope. Parameter $\sqrt{\delta_0}$ changes from
0 up to 10. We calculated polarization for $q=0;0.05;0.1;0.5$. For
values $\sqrt{\delta_0} > 10$ relatively good results can be
obtained by extrapolation from $p_{l}=100\,p_{\it
l}(\sqrt{\delta_0} =10)/\delta_0$ and for positional angles one
must be taken $\chi$ at boundary value $\sqrt{\delta_0}=10$ (see
figures).

The peak value of the polarization occurs when $\bf M$ is
perpendicular to the line of sight. In this case $\chi =0$, i.e.
the electric field of the integral polarization oscillates in the
plane ($\bf n M$). For $\delta_0 > 100$ and $q = 0$ we have the
asymptotic formula

\begin{equation}
p_{\it l}\cong \frac{6\%}{n\delta_0} \label{19}
\end{equation}

\noindent where $n=3,1,2$ corresponds to total dipole (14),
diamagnetic envelope (15) and Parker's outflow (16), respectively.
In figures these cases denoted as a), b) and c).

In the case $\vartheta_m =90^{\circ}$ the expressions (14), (15)
and (16) for $\delta\cos{\theta}$ differ only in different
coefficients before $\delta_0$ ($n = 3,1,2$, correspondingly). For
this reason, the peak polarization $p_{l\, max}$ in all cases is
the same but due to the different factors corresponds to different
values of $\sqrt{\delta_0}$. For the cases (14), (15) and (16),
$p_{l\, max}$ occurs at $\sqrt{\delta_0}$ to be equal to 1.385;
2.4 and 1.895, respectively. Of course, $p_{l\, max}$ depends on
the true absorption degree $q$. The calculations show that $p_{l\,
max}$ increases monotonically from 0.307\% at $q=0$ up to 1.01\%
at $q=0.5$.

The figures demonstrate that the position of $p_{l\, max}$ depends
weakly on the angle between $\bf M$ and line of sight $\bf n$. For
this reason, one can give, the common for all $\vartheta_m$ -
angles, approximate formula for wavelength $\lambda_{max}$ when
the peak polarization occurs. For $q=0$ the values
$\sqrt{\delta_0}$, mentioned above, give rise to the expression:

\begin{equation}
\lambda_{max}\approx \frac{2.68}{\sqrt{n\,B_0}} (\mu m),
\label{20}
\end{equation}

\noindent where $n=3,1,2$ corresponds to the models (14), (15) and
(16). Formula (20) allows us to estimate the magnetic field $B_0$
on the outer surface of spherical envelope if the value
$\lambda_{max}$ is found from polarimetric observations. For
$q\neq 0$ one can obtain the analogous formula.

The procedure is the following: from b) - curve one can find the
value $(\sqrt{\delta_0})_{max}$, divide it by $\sqrt{0.8}=0.894$
and take the number instead 2.86 in (20). For $q=0.5$ such
procedure gives 2.07 instead of 2.68, i.e. the existence of
absorption changes the estimation of $B_0$ only slightly. Roughly,
one can take $\lambda_{max}(\mu m)\approx B_0^{-1/2}(G)$.

Let us discuss now how to know from the polarimetric data whether
the envelope is optically thin or optically thick. First of all,
remember that the polarization degree of radiation, scattered in
optically thin envelope, as a rule is sufficiently higher than
that for optically thick envelope. But, the dilution effect from
the nonpolarized radiation can decrease this difference. For
example, the question arises from the analysis of AGN polarimetric
data how the dilution effect from the neighboring stars is
significant. It seems more informative to consider the spectral
differences of polarization in these cases. The optical depth of
the envelope, in principle, depends on wavelength. It can give the
jump of positional angle at some $\lambda$. This means the
transition from optically thin envelope to the optically thick one
(see the monograph [1]). The calculations in [8,9,11] for
optically thin envelopes are made only for pure dipole field (the
case (a) in our figures). Therefore, we shall compare only upper
figures with the curves in Fig.1.

The curves of the polarization degree $p_l$ in both types of
envelopes are qualitatively similar. First, the increasing of
$p_l\propto\lambda^4$ takes place, then the peak exists and
finally polarization degree decreases. This decreasing depends on
the explicit form of number density $N_e(r)$. For envelopes with
permanent concentration ($N_e=const$) we have
$p_l\propto\lambda^2$ at large values of $\lambda$. This
dependence coincides with the corresponding asymptotic behavior in
the case of optically thick envelope. But, most probably the
number density of free electrons falls with the removal from the
central source. If $N_e(r)$ is proportional to $r^{-2}$, then in
the case of optically thin envelope the polarization degree
decreases as $\propto 1 /\sqrt{\lambda}$ whereas optically thick
envelope gives $p_l\propto 1 /\lambda^2$.

Thus, the form of the polarization spectra at large wavelengths
allows to say whether the envelope is optically thin or thick.
Unfortunately, the value of polarization at large wavelength can
be very small (see (19)) and be undetectable from observations.
Besides, the asymptotic region of large wavelengths depends on
unknown magnetic field $B_0$ on the outer surface of the envelope.

To estimate the magnetic field $B_0$ for both types of envelopes
we need to know the spectrum of polarization near the $p_{l\,
max}$ value, i.e. we need to know $\lambda_{max}$. After that we
can use the relation (20) in the case of optically thick envelope.
Analogous expressions for optically thin envelopes are presented
in [8,9]. For most probable distribution $N_e\propto r^{-2}$ the
spectrum of polarization of optically thin envelope has the peak
at the following $\lambda_{max}$:

\begin{equation}
\lambda_{max}\approx \frac{4.65}{\sqrt{B_s\,\tau_{env}}} \simeq
\frac{4.65}{\sqrt{B_0\,\tau_{env}}}
\cdot\frac{R_s^{3/2}}{a^{3/2}}. \label{21}
\end{equation}

\noindent Here, $B_s= M/R_s^3$ is the dipole magnetic field at the
surface of central radiation source. The dipole law gives
$B_0=B_s\,R^3_s/a^3$. Because of $R_s <a$ and $\tau_{env} \leq 1$,
the estimation (21) gives rise to larger value for
$\lambda_{max}$, than the estimation (20). It should be noted that
for very large $B_0$ the value $\lambda_{max}$ can be in the X-ray
range of the wavelengths. If the value $B_0$ is assumed to be
known then the estimations from asymptotic formulae can be made
without knowledge of $\lambda_{max}$.

The spectra of the positional angles $\chi$ for optically thin and
thick envelopes are different. Generally speaking, this fact gives
an additional opportunity to distinct these both cases. However,
if the observed polarization corresponds to large wavelengths
(when $\delta_0 \gg 1$) then in both cases the angle $\chi$ has
practically permanent value and sign. In this case we can say
nothing to resolve the dilemma. Really, the spectrum of $\chi$ is
useful for this aim, when it is known near $\lambda_{max}$, where
the behavior of spectra for both cases is very different.

In the region of $\delta_0 \ll 1$ the $\chi$-spectrum is
practically useless as it seems that the signs of $\chi$ are
different in these cases. The sign of $\chi$ depends on the dipole
orientation which, as a rule, is unknown. If $\bf M$ changes
orientation to  opposite one, then $\chi$ changes its sign.
However, if from some reasons we know the $\bf M$ - orientation,
then, in contrary, the observed sign of $\chi$ demonstrate
directly what type of envelope exists - optically thin or
optically thick.

\section{SOME ASTROPHYSICAL APPLICATIONS}

\subsection{Polarization of radiation from OB and WR stars: the estimation of a magnetic field}

\begin{table}
\caption[]{The net polarization $p_l$ of hot stars for various
magnitudes of equatorial magnetic field.}
\begin{flushleft}
\begin{tabular}{||l|l|l|l||}
\hline\hline Objects  &~~$\tau_{T}$ &~~$B_{l}=100$G &~~$B_{l}=10$G
\\ \hline\hline $\zeta$ Puppis &~~0.2 &~~$p_{l}=0.4\%$ &~~0.02\%
\\ \hline $\varepsilon$ Ori &~~0.17 &~~0.17\% &~~0.01\% \\ \hline
Deneb &~~0.03 &~~0.01\% &~~$\ll$ 0.01\% \\ \hline P Cyg &~~1
&~~0.1\% &~~0.3\% \\ \hline WR40 &~~3.4 &~~0.1\% &~~0.3\% \\
\hline\hline
\end{tabular}
\end{flushleft}
\end{table}

The hot stars are characterized by very strong, $\sim\dot{M}\geq
10^{-5}M_{\odot}$, outflow of a matter in form of the dense
stellar wind (see [16-18]). The Thomson optical depth $\tau$ of
such wind can be both less ($\zeta$ Puppis, Deneb) and greater (of
the order) than unity (P Cyg, WR 40)(see [17]). The problem of the
determination of surface magnetic field for these objects exists
for a long time. In the paper [19] the positive answer is given
for the question whether dipole magnetic field can exist at the
surface of these stars.

The attempts to measure the magnetic fields of bright stars by the
traditional circular polarization technique were unsuccessful
[16,20]. Particulary, in the paper [20] the upper limit of the
magnetic field component along the line of sight was obtained for
the star 04I(n)f $\zeta$ Puppis: $B_{||}\leq 200 G$. The
sufficient variability of the Stokes parameter $V$ was not also
found. One can expect that future Spectro-Polarimetric
Interferometer (SPIN) will be useful.

Using our calculations (Fig.1 and 2) one can estimate the
polarization in V-band for a number of stars from the list of
objects of the project SPIN [17]. So, assuming that equatorial
magnetic field $B_{e}\approx 10^{2}$G we obtain $\delta_{0}\approx
0.8 \lambda_{V}^{2} B_{e}\approx 24.2$. If $B_{e}\approx 10$ then
we have $\delta_{0}\approx 2.42$. The values of expected
polarization degree are presented in Table 3 (for the inclination
angle $i=90^{\circ}$). In estimations we have chosen the wind
density distribution from the paper [21].

\subsection{Polarization of radiation from the system Cyg X-1/HDE 226868}

In 1975 Nolt et al. [22] have discovered the intrinsic linear
polarization of the optical radiation of close binary system Cyg
X-1/HDE 226868 with a compact component being a black hole with
the mass $\sim 10M_{\odot}$. The polarization has variations with
the amplitude $\sim 0.2$\%. The time behavior of polarization is
rather complex. Besides the period of the binary system $P=5^d.6$,
the polarization varies with $39^d$ and $78^d$ characteristic
times [22]. The possible mechanisms of the polarization of this
system were discussed in [26].

If the polarized radiation arises in the accreting matter around
the black hole the value of intrinsic polarization is to be some
tens times higher than the observed value. It seems there exists
very strong dilution by the unpolarized emission of supergiant HDE
226868. The luminosity of this star is estimated as $L_{0}\approx
(1\div 3)\times 10^{39}\,erg\, s^{-1}$ whereas the X-ray
luminosity of Cyg X-1 is about $L_{X}\leq 8\times 10^{37}\, erg\,
s^{-1}$ [23]. If optical radiation of the accretion disk arises as
a result of the transformation of X-ray emission of the black hole
then the value of intrinsic polarization have to be near the value
$p_{l}\leq 10\%$. Optically thick accretion disk with most
probable inclination $i=30^{\circ}$ can not produce such high
polarization. For this reason if the polarization originates into
accreting matter, then this polarization can arise only as a
result of the scattering in optically thin coronae or in some sort
of outflow from the accretion disk (wind). At the same time, it is
not possible to exclude the mechanism of the production of small
(at the level of some tens percents) intrinsic polarization from
the star HDE 226868 in a magnetized stellar wind.

We consider both mechanisms and the corresponding consequences.

In the case of optically thin magnetized corona around the
accretion disk the large, $\sim p_l\sim 10\%$, polarization can be
produced if the central source is point-like or the coronae itself
is very vast. For such model one can use the calculations of
Dolginov et al. [1]. According to their calculations $p_l\sim
10\%$ can be obtained at $\delta_0\tau_{env}\approx 10$ and when
magnetic field is perpendicular to the line of sight. For
$i=30^{\circ}$ it means that the magnetic field lies almost in the
plane of the disk. If one supposes $\tau_{env}\approx 0.1$ then
the condition $\delta_0\approx 100$ (remember that
$\delta_0\tau_{env}\approx 10$) gives rise to the estimation
$B_0\approx 500\, G$. In a number of models the inner radius of a
hot coronae is determined as $\sim 100\, R_g$, where $R_g=25
M/c^2$ is the gravitational radius. In this case for dipole
magnetic field follows the estimation $B(3R_g)\sim 10^7\, G$.

Let us now consider the situation when the linear polarization is
generated in the extended stellar wind of the optical component
HDE 226868. The existence of such wind was confirmed by
observation of modulated radio emission from Cyg X-1. The radio
emission of the relativistic jet of the black hole has variable
absorption in the plasma of the stellar wind. Modulation arises in
the process of the orbital motion of the black hole [27].

Supposing that in stellar wind $N_e\sim r^{-2}$ and
$\dot{M}\approx 2\times 10^{-6} M_{\odot}\, year^{-1},\,
R_s=10R_{\odot}$ and $V_{\infty}\approx 1850\, km\, s^{-1}$ [27],
one can estimate the Thomson optical depth of the wind as
$\tau_T\approx 0.1$. According to Fig.1 the polarization
$p_l\approx 0.2\%$ corresponds to the angle between magnetic
dipole $\bf M$ and line of sight being $\sim 65^{\circ}$ and the
parameter $\sqrt{\delta_S\tau_{env}}\approx 3$. It gives the
estimation $B_s\approx 350\, G$. Such value of extremal magnetic
field for the OV supergiant seems to be rather large. It is
interesting that in this case the dipole axis lies in the orbital
plane and one has not any problems to explain the variability of
the polarization.

\subsection{Intrinsic polarization of SS 433}

The famous galactic X-ray binary system SS 433 is the microquasar
with two relativistic ($V\approx 0.26c$) strongly collimated
($\vartheta\approx 1^{\circ}$) jets. The system is characterized
by 3 periodic motions: the orbital motion with the period
$P_o=13^d.082$, the precession with $P_p=162^d.5$ and nutation
with $P_n=6^d.28$ [23,28]. Recent spectral observations [20]
confirm the existence of spectral absorption lines of the A7Ib
supergiant. The mass ratio of the relativistic component and an
optical star is about $q=M_{X}/M_{V}=0.57\pm 0.11$. The masses of
every component are estimated as $M_{V}=(19\pm 7)M_{\odot}$,
$M_{X}=(11\pm 5)M_{\odot}$ [28]. The kinematics of relativistic
jets allows to estimate the inclination angle of the orbital plane
$i=78^{\circ}.82\pm 0^{\circ}.11$. The precession angle $\theta$
of a disk is equal to $\theta =19^{\circ}.80\pm 0^{\circ}.18$.

The variable, i.e. intrinsic, polarization of SS 433 first was
observed by McLean and Tapia [30], and later, in B, V, R, I bands,
by the observers in KrAO [31]. In 1993 the polarimetric
observations of SS433 were obtained in UV range by high speed
photometer-polarimeter (HSP) from the cosmic Hubble telescope
[32]. The polarization degree in UV range is rather high
$p_{l}=(13.4\pm 4)\%$ and it demonstrates the time variability.
Such a high degree of polarization can not originate in an
accretion disk. Most probably it is a result of the scattering of
radiation by free electrons in a plasma outflow (for example, in a
hot corona or wind). An example of such plasma can be, also,
optically thin advective accretion flux around the black hole
[33].

We calculated the expected polarization degree of the radiation
scattered in a spherical stellar wind with the Parker's type
magnetic field

\begin{equation}
B_{r}=B_{p}(R_{A}/r)^{2}\cos{\theta};\,\,\,\,
B_{\varphi}=B_{T}\frac{R_{A}}{r}\sin{\theta};\,\,\,\, B_{\theta}=0
\label{22}
\end{equation}

\noindent where $B_p$ is the magnetic field at magnetosphere's
pole, $B_T$ is the value of toroidal component, $R_A$ is the
radius of the magnetosphere. There was used the technique
described in the papers [1,8-10].

The results are presented in Fig.6. They are compared with the
observational data of Dolan et al. [32]. We assumed that the
Thomson optical depth of the coronae is near unity, i.e.
$\tau_T\leq 1$.

The results of UV polarimetric observations corresponds rather
well to the theoretical curve if one assumed $B_p=1\, G$ and
$B_T=600\, G$. The disagreement between observational polarization
and the theory in optical range can be explained by the existence
of the dilution from intrinsic non-polarized radiation of the
accretion disk. However, the Thomson polarization of radiation
outgoing from plane-parallel disk at $i\approx 80^{\circ}$ is
rather high ($p_l\approx 5\%$, [13]), the strong ($B_T\approx
600\, G$) toroidal magnetic field depolarizes fully the scattered
radiation because of the depolarization parameter $\delta_0$ is
very large ($\delta_0\geq 500(\lambda/1\mu m)^2$, see (21)).

It is interesting that our estimation of the magnetic field near
the magnetosphere allows us to obtain the additional information
about the parameters of the accretion to the black hole itself.

The modern theory of rotating black holes determines the
magnetosphere as a region from which the relativistic jets outflow
occurs and, simultaneously, arises the hot coronae [4,34].
According to [34], the radius $R_A\approx 5\times 10^3R_g$ in the
case of pure thermal coronae and $R_A\approx 300R_g$ for the, so
called, hybrid coronae. As a result one can estimate the value of
magnetic field in the vicinity of black hole, near the last stable
orbit $3R_g$ ($R_g$ is the gravitational radius). If one assumed
that the estimation (21) for Parker's type of magnetic field is
valid up to $3R_g$, then we obtain the estimation of the magnetic
field of the black hole in SS 433 as $B_{max}\approx 10^4\div
10^6\, G$, depending on the type of coronae. This value is in a
good agreement with that in the model of Blandford and Znajek [4].

It should be noted that if one takes $R_A\approx 10^{10}\, cm$ (as
in [34]) then the value of magnetic moment $\mu$ of the compact
component in SS 433 can be estimated as $\mu\approx 10^{34}\, G\,
cm^{3}$. This value is near the value of $\mu$ for another
microquasar GRS1915+05 obtained by Robertson and Leiter [2] from
very different consideration.

Finally, if, according to [33], one considers that the magnetic
force lines strongly follow the motion of the plasma in the
accretion disk, then we can estimate the known viscosity factor
$\alpha$ of the Sunyaev-Shakura theory [35]. According to

\begin{equation}
\frac{B_{r}}{B_{\varphi}}=\frac{U_{r}}{U_{\varphi}}\approx\alpha
\label{23}
\end{equation}

\noindent where $U_r$ and $U_{\varphi}$ are the radial and
Keplerian velocities in accretion disk, correspondingly. Our
estimation of $B_r$ and $B_{\varphi}$ gives rise to $\alpha\approx
2\times 10^{-3}$ what is practically coincides with the
theoretical estimation in [36].

\subsection{Polarization of radiation of supernovae}

Recently a lot of observational data has been obtained giving an
evidence that the radiation from supernovae has a sufficient
polarization (see the review [37]). In particular, the
spectropolarimetric observations of young supernovae, obtained at
Kek-telescope, demonstrate that the intrinsic polarization exists
for all the types of supernovae [38]. These observations, for
example, show that the supernovae 1997dt (Ia), 1998T and 1997dq
(both of the type Ib), 1997ef (Ib/c - pecular), 1997eq (IIn),
1997ds (II-p) have intrinsic polarization. It was shown [39] that
the supernovae of types II and Ib/c are polarized at the level
$p_l\approx 1\%$. Some of them have higher polarization. It is
interesting (see [39]) that the lesser is the hydrogen envelope
and the deeper layers are seen the larger is the value of
polarization. Such dependence was analyzed by Hofflich et al.
[40]. It proved to be the case that the observed polarization is
higher for SN II as compared with SN Ic.

It is natural, that many authors explain the observed polarization
as due to the scattering of radiation in asymmetric envelope of
the expulsed matter (see, for example, [37]).

There exists another physical mechanism that can be responsible
for the noticeable net polarization of SN radiation without
asymmetric distribution of the exploded matter.

Such mechanism is the considered above integral effect of the
Faraday rotation of the polarization plane of the radiation
scattered in the spherically symmetric envelope of exploded matter
(blow shock) with a magnetic field. Namely this effect produces
the intrinsic polarization. The measured net polarization allows
to estimate the magnetic field strength in the region of
propagating blow shock.

Now let consider some examples. The supernova SN IIp 1999em was
observed at epochs of 7, 40, 49, 159 and 163 days after the
explosion and its radiation was to be polarized at these epochs
(Leonard et al., 2001). The detected net polarization was
$p_{l}\sim (0.2\div 0.5)\%$. This magnitude is quite well
corresponding to the results of our calculations presented at
Fig.2, the polarization maximum corresponding to the Faraday
depolarization parameter $\delta_{0}\sim 4$. For the visual
magnitude (V band - $\lambda_{eff}=0.55\mu m$) this value of the
depolarization parameter allows to determine the magnetic field
strength as $B\approx 16.5\, G$.

Another example is SN Ia. The typical value of the intrinsic
polarization of such kind supernovae is to be $p_{l}\approx 0.7\%$
(Kasen et al., 2003). This value of polarization is better
corresponding to our results presented at Fig.1, i.e. to the case
of scattering into the magnetized optically thin envelope. In this
case the real depolarization parameter is
$(\delta_S\tau_T)^{1/2}$, where $\tau_T \ll 1$ is the optical
thickness of an envelope with respect to the electron scattering.
The polarization magnitude $p_{l}\sim 1\%$ corresponds to the
depolarization parameter value of $(\delta_S\tau_T)\sim 1$. In
this case the magnetic field strength is $B_s\sim 4/\tau_T\, G$.
For $\tau_T\sim 0.1$ the magnetic field strength in the shock
region can reach the strength $\sim 40\, G$ and even $\sim 100\,
G$.

The calculation of non-sphericity of an envelope is not principal
problem. Some cases of generation of polarized radiation in a
magnetized non-spherical envelope was considered in [1,14].

\section{CONCLUSIONS AND OUTLOOK}

We represented the results of calculations of the integral
polarization of radiation emerging from the optically thick
spherically symmetric envelope with the dipole magnetic field of a
central radiation source (a neutron star; extended supernova
envelopes produced by the supernova explosion at the initial stage
of expansion; accretion flows around quasars and active galaxy
nuclei). We considered also the cases of diamagnetic (perfectly
conducting) plasma envelope when the normal component of magnetic
field does not penetrate inside a disk and radially evaporated
envelope. The comparison of the results of polarization between
the various types of envelopes have been made.

The results of our theoretical calculations can be used for the
analysis of polarimetric observations of hot stars, including WR
stars, X-ray binaries (Cyg X-1/HDE 226868, SS 433), supernovae and
etc. We plan to use the results of our calculations to an analysis
of polarimetric data of quasars and AGNs (see, for example, [44]).

This work is partially supported by the RFBR grant 03-02-17223,
the Program of the Prezidium of RAN "Nonstationary Phenomenae in
Astronomy", the Program of the Department of Physical Sciences of
RAN "The Extended Structures..." and by the Program of Astronomy
of Russian Science and Education Ministry.

\begin{figure*}
\includegraphics[width=16cm]{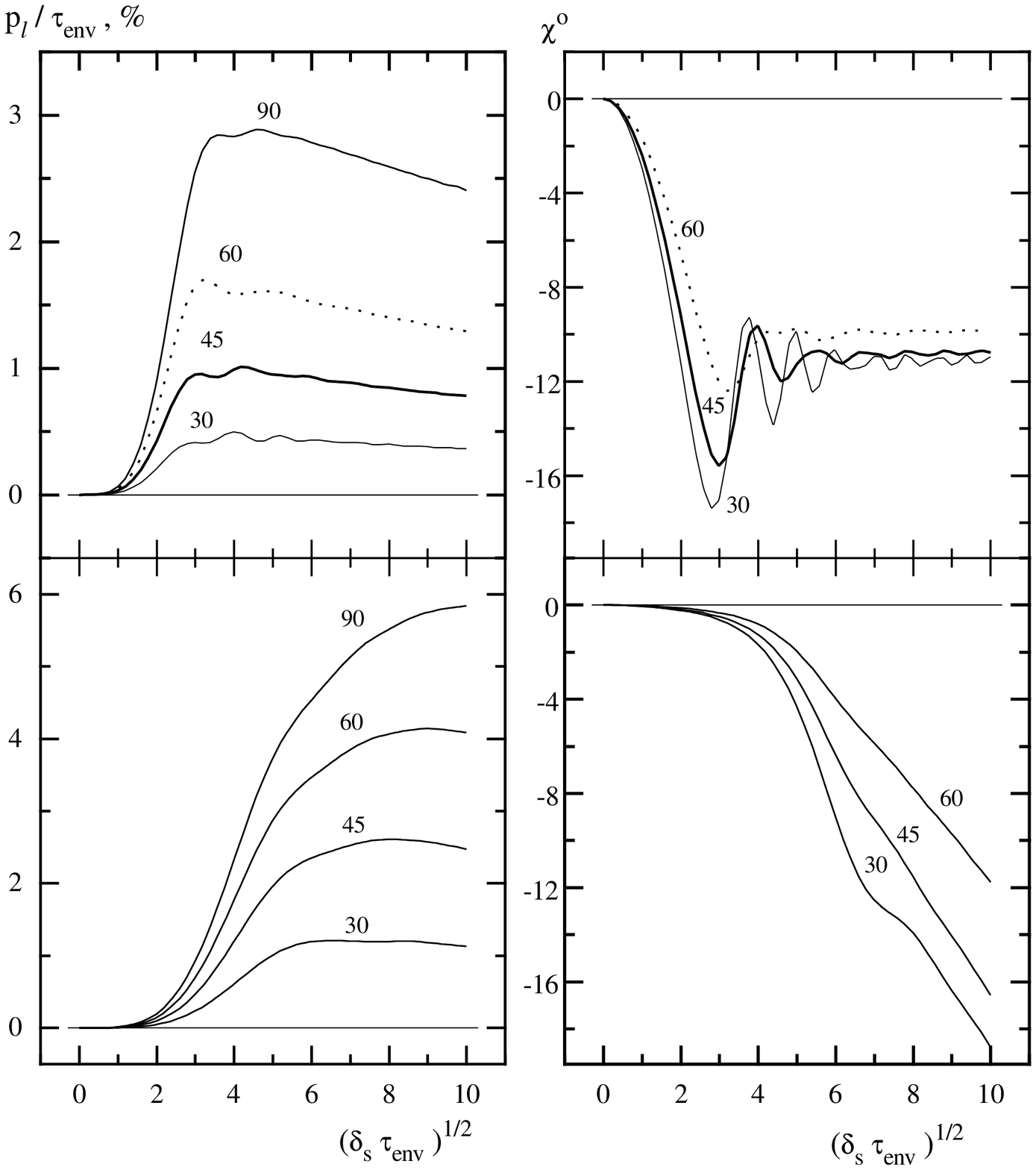}
\caption{The polarization degree and the positional angle of the
integral radiation of optically thin spherical envelope in the
magnetic dipole field. The upper panels corresponds to the case
$N_e\sim r^{-2}$ and the low ones to the envelope with constant
electron number density $N_e = const$ with $\eta =5$. The numbers
near the curves denote the angle (in degrees) between the magnetic
dipole $\bf M$ and line of sight $\bf n$.}
\end{figure*}

\begin{figure*}
\includegraphics[width=16cm]{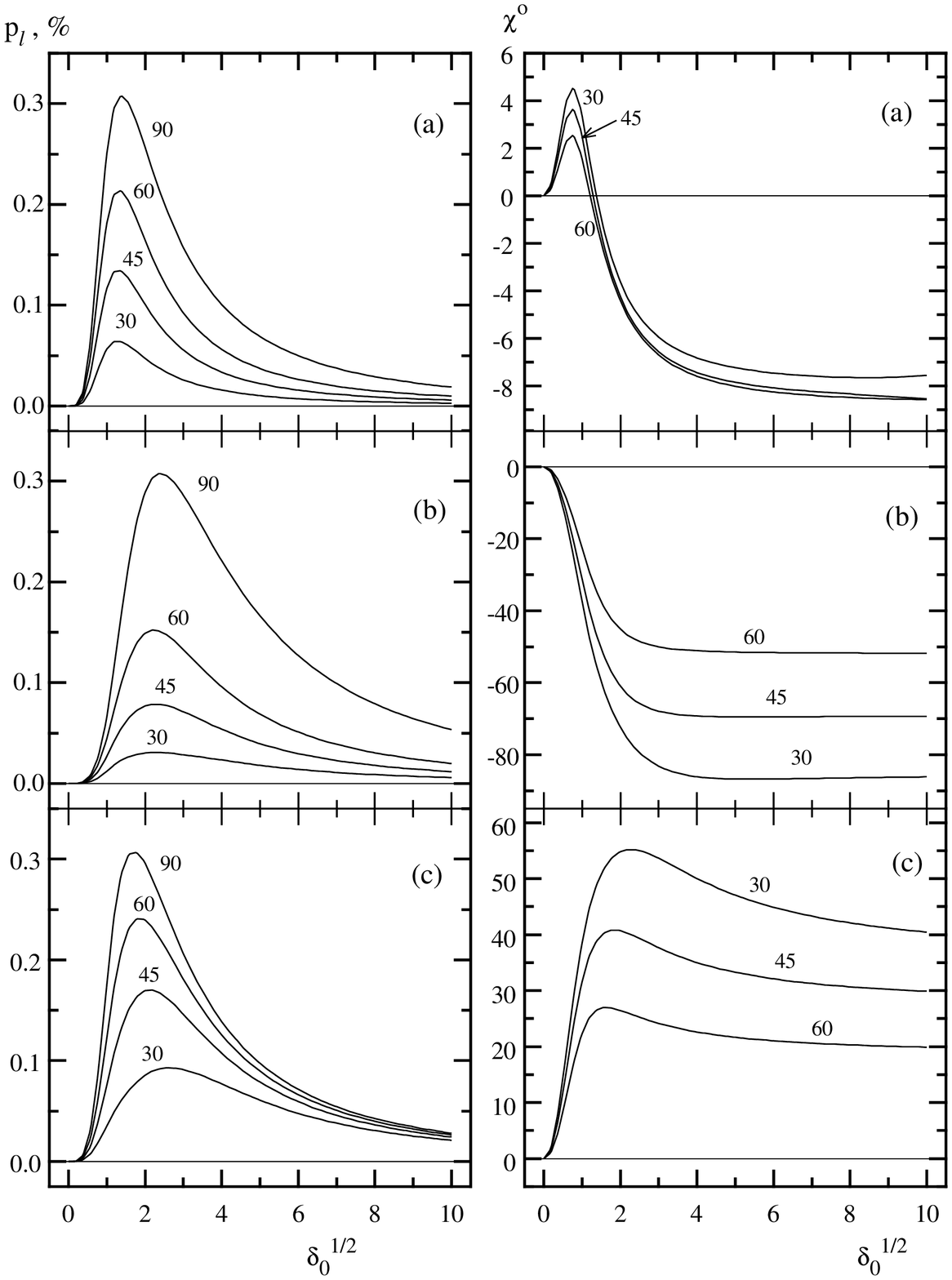}
\caption{The polarization degree and the positional angle of the
integral radiation of the optically thick magnetized for the
conservative atmosphere ($q=0$). The numbers denote the angle
$\vartheta_m$ (in degrees) between the magnetic dipole $\bf M$ and
the line of sight $\bf n$. The cases a), b) and c) correspond to
the total dipole field, diamagnetic atmosphere ($B_r =0$) and
Parker's outflow field ($B_r \neq 0$), correspondingly.}
\end{figure*}

\begin{figure*}
\includegraphics[width=16cm]{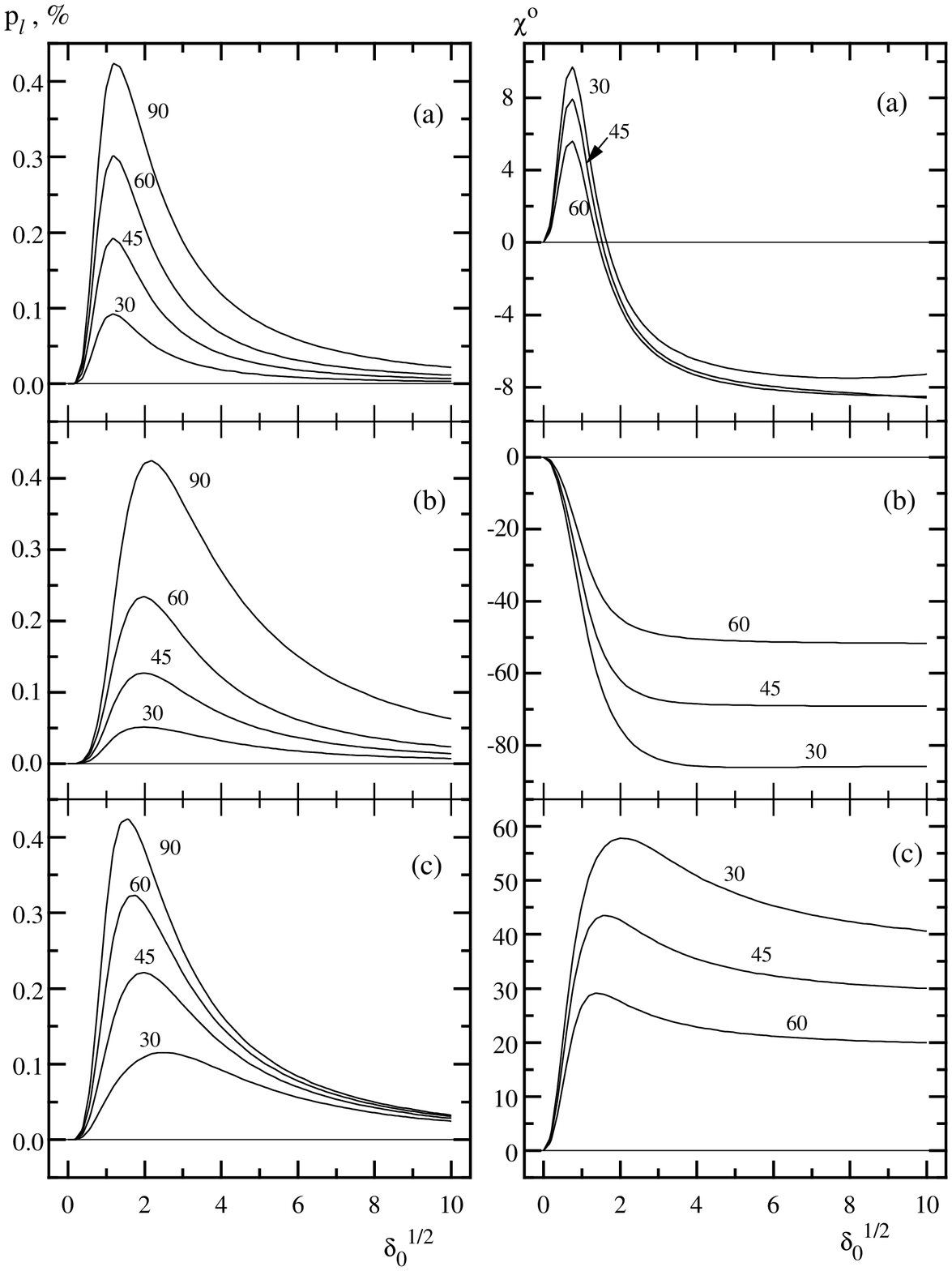}
\caption{The same as in Fig.2, but for $q=0.05$.}
\end{figure*}

\begin{figure*}
\includegraphics[width=16cm]{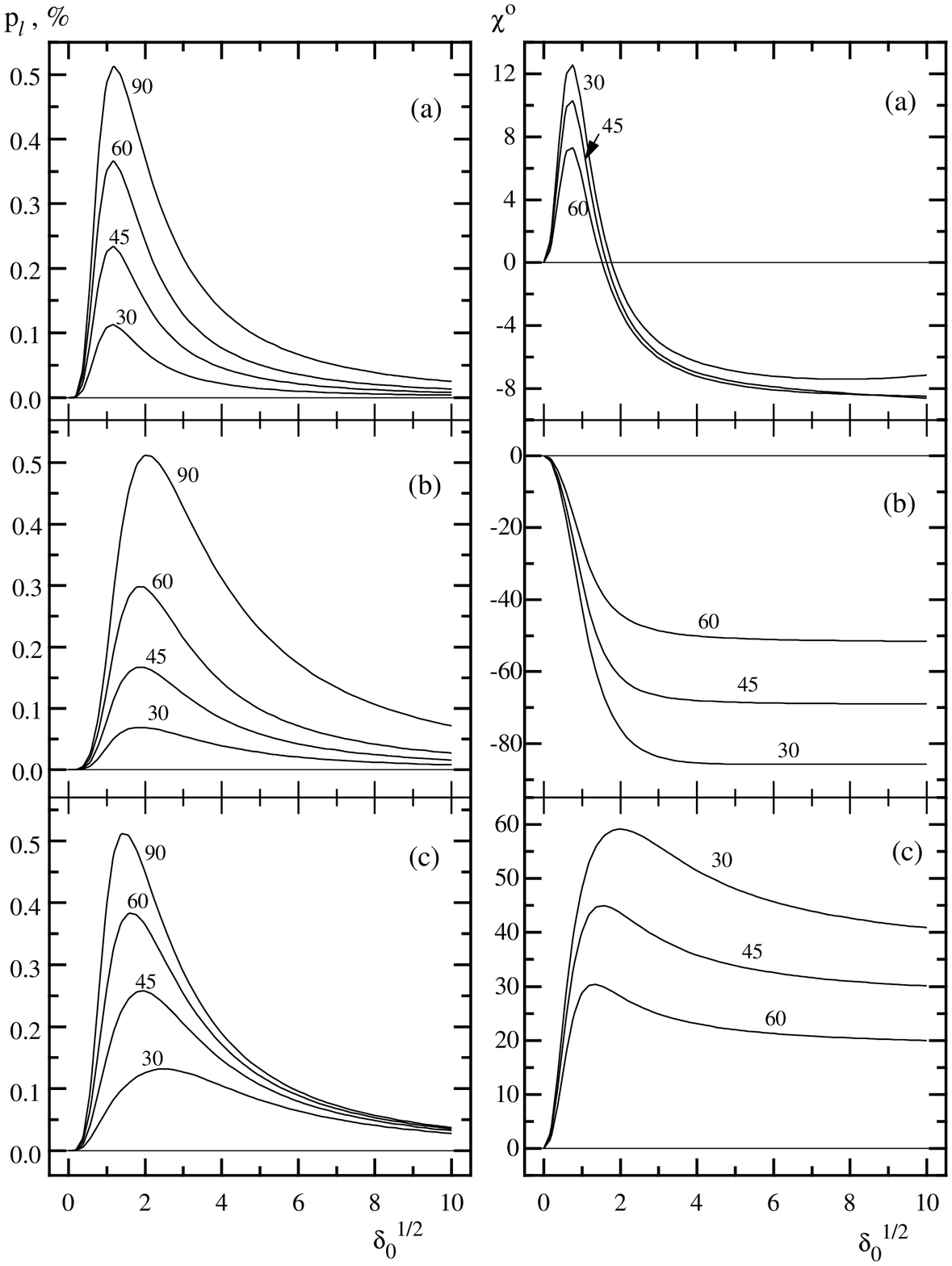}
\caption{The same as in Fig.2, but for $q=0.1$.}
\end{figure*}

\begin{figure*}
\includegraphics[width=16cm]{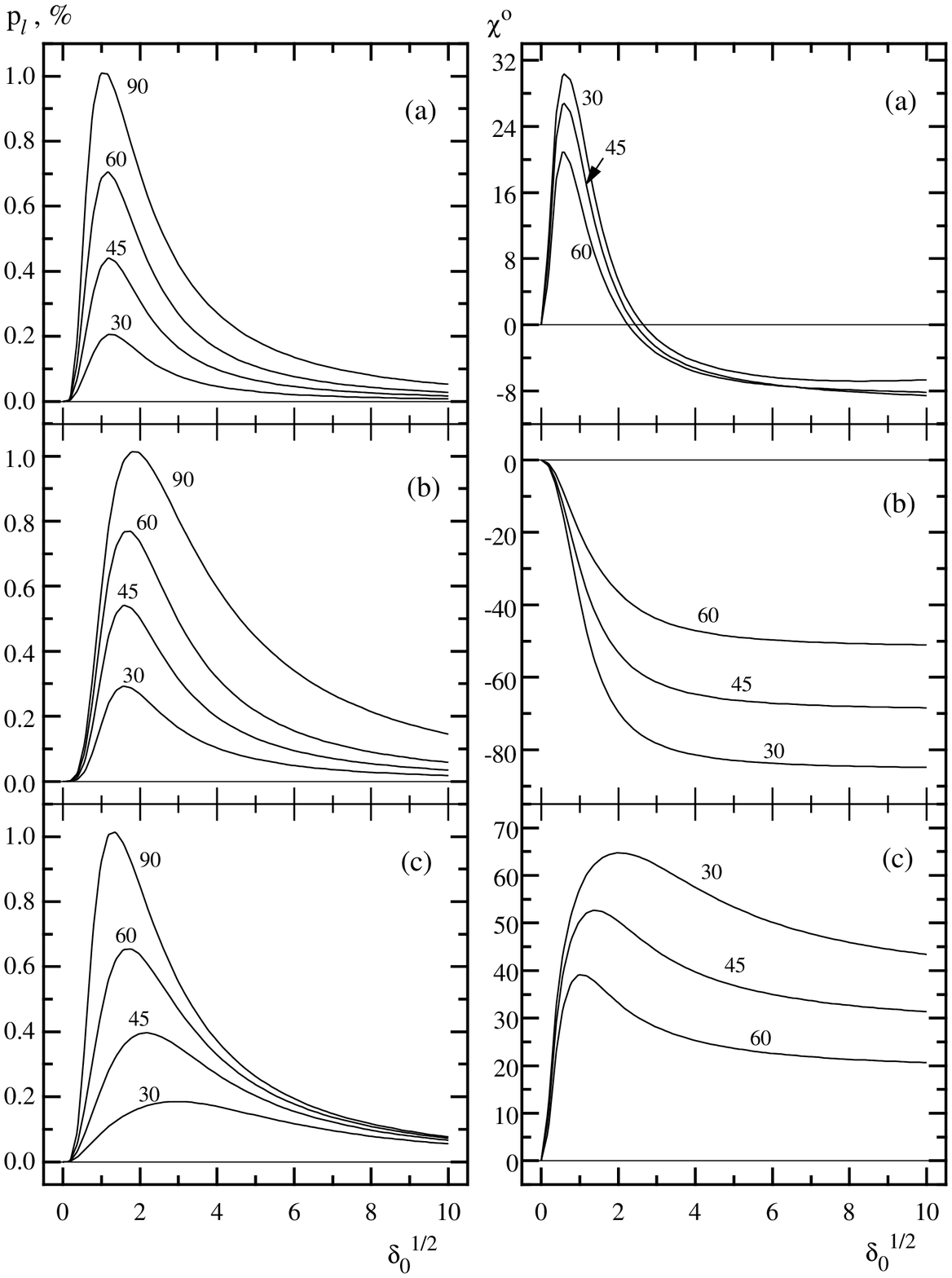}
\caption{The same as in Fig.2, but for $q=0.5$.}
\end{figure*}

\begin{figure*}
\includegraphics[width=16cm]{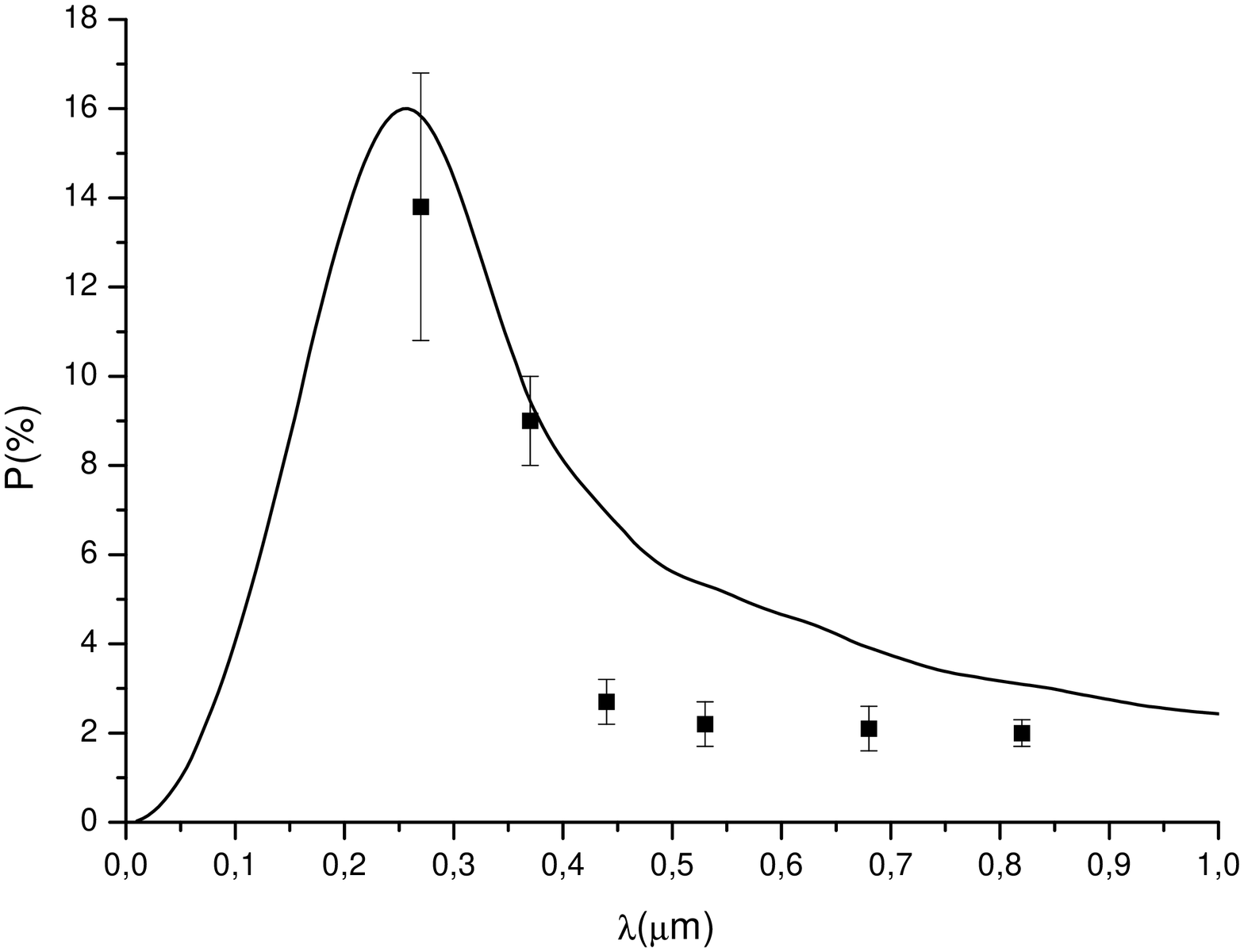}
\caption{The spectrum of integral linear polarization of X-ray
binary system SS433: comparison of the observational with the
results of theoretical model (solid curve) at $\tau_T =1.0$,
$B_{Pol} =1G$, $B_{Tor} =600G$.}
\end{figure*}


\begin{thebibliography}{99}
\bibitem{1}A. Z. Dolginov, Yu. N. Gnedin, N. A. Silant'ev,
{\it Propagation and polarization of radiation in cosmic media}
(Gordon and Breach Publs., Amsterdam, 1995)
\bibitem{2}S. L. Robertson and D. J. Leiter, Astrophys.J., {\bf
596}, October 20 (2003).
\bibitem{3}Yu. N. Gnedin, N. V. Borisov, T. M. Natsvlishvili, M. Yu.
Piotrovich, astro-ph/0304158.
\bibitem{4}R. D. Blandford and R. L. Znajek, MNRAS, {\bf 179}, 433 (1977).
\bibitem{5}L. -X. Li, astro-ph/0202361.
\bibitem{6}F. M. Rieger and K. Manheim, astro-ph/0011012.
\bibitem{7}A. Tomimatsu and M. Takakashi, Astrophys.J., {\bf 552}, 710 (2001).
\bibitem{8}Yu. N. Gnedin and N. A. Silant'ev, Pisma AZh, {\bf 6},
344 (1980).
\bibitem{9}Yu. N. Gnedin and N. A. Silant'ev, Asrophys.Sp.Sci., {\bf
102}, 375 (1984).
\bibitem{10}N. A. Silant'ev, Yu. N. Gnedin, and T. Sh. Krymski,
Astron.Astrophys., {\bf 357}, 1151 (2000).
\bibitem{11}M. A. Pogodin, Pisma AZh, {\bf 18}, 442 (1992).
\bibitem{12}E. Agol, O. Blaes, and C. Ionescu-Zanetti, MNRAS,
{\bf 293}, 1 (1998).
\bibitem{13}P. S. Shternin, Yu. N. Gnedin and N. A. Silant'ev,
Astrofizika, {\bf 46}, 433 (2003).
\bibitem{14}N. A. Silant'ev, Astron.Astrophysics, {\bf 383}, 326 (2002).
\bibitem{15}D. Lai, Astrophys.J., {\bf 524}, 1030 (1999).
\bibitem{16}J. Babel and T. Montmerle, Astrophys.J.Lett., {\bf 485}, L29 (1997).
\bibitem{17}O. Chesneau, S. Wolf, and A. Domiciano de Souza,
astro-ph/0307407.
\bibitem{18}M. E. Contreras, G. Montes and F. P. Wilkin,
astro-ph/0310393.
\bibitem{19}A. B. Underhill and R. P. Fahey, Astrophys.J., {\bf 280}, 712 (1984).
\bibitem{20}O. Chesneau and A. F. J. Moffat, PASP, {\bf 114}, 612 (2002).
\bibitem{21}J.P. Cassinelli, N.M. Hoffman, MNRAS, {\bf 173}, 789
(1975).
\bibitem{22}I.G. Nolt, J.C. Kemp, R.J. Rudy et al., Ap.J.L., {\bf
199}, L27 (1975).
\bibitem{23}A. M. Cherepashchuk, Uspehi Fiz. Nauk, {\bf 171}, 864
(2001).
\bibitem{24}N. G. Bochkarev, E. A. Karitskaya, R. A. Sunyaev, N. I. Shakura, Astron.Zh., {\bf 55}, 185 (1979).
\bibitem{25}E. A. Karitskaya, Astron.Zh., {\bf 58}, 146 (1981).
\bibitem{26}Yu. N. Gnedin, N. V. Borisov, T. M. Natsvlishvili, M. Yu. Piotrovich, N. A. Silant'ev, astro-ph/0304158, Astrophys.Sp.Sci.,
in press (2004).
\bibitem{27}C. Brocksopp, R. P. Fender and G. G. Pooley,
astro-ph/0206460, MNRAS in press (2003).
\bibitem{28}A. M. Cherepashchuk, R. A. Sunyaev, E. V. Seifina, I. E. Panchenko, S. V. Molkov, K. A. Postnov, astro-ph/0309140 (2003).
\bibitem{29}D. R. Gies, W. Huang and M. V. McSwain, Astrophys.J., {\bf 579}, 67 (2002).
\bibitem{30}I. S. McLean and S. Tapia, Nature, {\bf 287}, 704 (1980).
\bibitem{31}Y. S. Efimov, V. Piirola and N. M. Shakhovskoy,
Astron.Astrophys., {\bf 138}, 62 (1984).
\bibitem{32}J. F. Dolan, P. T. Boyd, S. N. Fabrika, {\it et al.}
Astron.Astrophys., {\bf 327}, 648 (1997).
\bibitem{33}L. -X. Li, astro-ph/0112503 (2002).
\bibitem{34}T. J. Maccarone and P. S. Coppi, astro-ph/0204235 (2002).
\bibitem{35}N. I. Shakura and R. A. Sunayev, Astron.Astrophys., {\bf 24},
377 (1973).
\bibitem{36}A. R. King, J. E. Pringle, R. G. West, M. Livio, astro-ph/0311035 (2003).
\bibitem{37}L. Wang, D. Baade, P. H\"{o}flich, J. C. Wheeler,  The
Messenger, {\bf 109}, 47 (2002).
\bibitem{38}D. C. Leonard, A. V. Fillipenko, A. J. Barth, T. Matheson, Astrophys.J., {\bf 536}, 239 (2000).
\bibitem{39}J. C. Wheeler, P. H\"{o}flich, L. Wang, I. Yi,
astro-ph/9912080 (1999).
\bibitem{40}P. H\"{o}flich, J. C. Wheller, and L. Wang, Astrophys.J., {\bf 521}, 179 (1999).
\bibitem{41}D. C. Leonard, A. V. Fillipenko, and M. S. Brotherton, Astrophys.J., {\bf 553}, 861 (2001).
\bibitem{42}D. Kasen, P. Nugent, L. Wang, {\it et al.} astro-ph/0301312 (2003).
\bibitem{43}P. G. Martin, I. B. Thompson, J. Naza, J. R. P. Angel, Astrophys.J., {\bf 265}, March 15 (1983).
\bibitem{44}Yu. N. Gnedin and N. A. Silant'ev, Pisma AZh, {\bf
28}, 490 (2002).
\end{thebibliography}
\end{document}